\mag=1200
\input amstex
\documentstyle{amsppt}
\NoBlackBoxes
\TagsOnRight
\tolerance=1000
\voffset=-12mm

\let\le=\leqslant
\let\ge=\geqslant

\let\veps=\varepsilon
\let\myp=\partial
\let\myl=\overline
\let\myk=\varkappa
\let\pfi=\varphi
\let\myt=\times
\let\myw=\widetilde
\let\mys=\subset
\define\INT{\int_{-\infty}^\infty}
\define\RR{\Bbb R}
\define\CC{\Bbb C}
\define\erf{\operatorname{erf}}
\define\sgn{\operatorname{sgn}}
\redefine\Re{\operatorname{Re}}
\topmatter 
\address
Steklov Mathematical Institute,\hfil\break\indent
Russian Academy of Sciences
\endaddress
\email
vladim\@mi.ras.ru
\endemail
\author
V.~S.~Vladimirov
\endauthor
\title
On the equation of the $p\,$-adic open string 
for the scalar tachyon field
\endtitle
\abstract We study the structure of solutions of the one-dimensional
non-linear pseudodifferential equation describing the dynamics of the
$p$-adic open string for the scalar tachyon field
$p^{\frac12\partial^2_t}\Phi=\Phi^p$. We elicit the role of real zeros of
the entire function~$\Phi^p(z)$ and the behaviour of solutions~$\Phi(t)$
in the neighbourhood of these zeros. We point out that discontinuous
solutions can appear if~$p$ is even. We use the method of expanding the
solution~$\Phi$ and the function~$\Phi^p$ in the Hermite polynomials and
modified Hermite polynomials and establish a connection between the
coefficients of these expansions (integral conservation laws). For $p=2$
we construct an infinite system of non-linear equations in the unknown
Hermite coefficients and study its structure. We consider the
3-approximation. We indicate a connection between the problems stated and
the non-linear boundary-value problem for the heat equation.

\endabstract
\endtopmatter
\rightheadtext{On the equation of the $p$-adic open string}
\footnote""{This research was supported in part by the grant of the
President of Russian Federation (grant no.~NSh-1542.2003.1) and the
Russian Science Support Foundation.}
\document

\head \S\,1. Introduction
\endhead

The dynamics of the open~$p$--adic string for the scalar tachyon field is
described by the non-linear pseudodifferential equation~\cite{1}--\cite{9}
$$
p^{\frac12\square}\Phi=\Phi^p,
\tag1.1
$$
where
$$
\square=\myp^2_t-\myp^2_{x_1}-\cdots-\myp^2_{x_{d-1}}, \qquad
t=x_0,
$$
is the d'Alembert operator and~$p$~is a prime number, $p=2,3,5,\dots$\,.
In what follows~$p$ is any positive integer. We consider only real
solutions of equation~\thetag{1.1}, since only real solutions have
physical meaning.

In the one-dimensional case ($d=1$) we use the change
$$
\pfi(t)=\Phi\bigl(t\sqrt{2\ln p}\,\bigr)
$$
and write equation~\thetag{1.1} in the following equivalent form:
$$
e^{\frac12\myp^2_t}\pfi=\pfi^p.
\tag1.2
$$

Equation~\thetag{1.2} is a non-linear integral equation of the following
form~\cite{9}:
$$
\frac1{\sqrt{\pi}}\int^\infty_{-\infty}e^{-(t-\tau)^2}\pfi(\tau)\,d\tau
=\pfi^p(t), \qquad
t\in\RR.
\tag1.3
$$

\goodbreak
Solutions of equation~\thetag{1.3} are sought in the class of measurable
functions~$\pfi(t)$ such that
$$
|\pfi(t)|\le C\exp\{(1-\veps)t^2\} \quad \text{for any} \quad \veps>0,
\quad t\in\RR. \tag1.4
$$

The following boundary-value problems for the solutions $\pfi$ of
equation~\thetag{1.3} have physical meaning:
$$
\lim_{t\to-\infty}\pfi(t)=0, \qquad \lim_{t\to\infty}\pfi(t)=1 \tag1.5
$$
if~$p$ is even, and
$$
\lim_{t\to-\infty}\pfi(t)=-1, \qquad \lim_{t\to\infty}\pfi(t)=1 \tag1.6
$$
if $p$ is odd.

In \S\,2 we give the information on Hermite polynomials~$H_n(t)$, \
$n=0,1,2,\dots$, that will be used in the subsequent sections, introduce
modified Hermite polynomials~$V_n(t)$, \ $n=0,1,\dots$, and study their
properties. In~\S\,3 we study the properties of the integral operator~$K$
on the left side of equation~\thetag{1.3}, in the space~$L_2^\alpha$, \
$0<\alpha<2$. In~\S\,4 we expand the solution~$\pfi(t)$ of
equation~\thetag{1.3} in series in polynomials~$H_n$ and~$V_n$,
expand~$\pfi^p(t)$~in Taylor series and in series in~$H_n$, and establish
relations between these expansions. In particular, we prove an important
equality~\thetag{4.5}, which plays the role of integral conservation laws:
$$
(\pfi^p,H_n)_1=(\pfi,V_n)_{1/2}, \qquad
n=0,1,\dots\,.
$$

In \S\,5 we suggest a new method of study of equation~\thetag{1.3}, which
enables us to reduce the problem of solving this equation to the problem
of solving a non-linear boundary-value problem for the heat equation. We
use this equivalence to prove Theorem~5.1. In~\S\,6 we consider the linear
equation~\thetag{1.3} ($p=1$), establish a connection between its
solutions and the solutions of the heat equation~\thetag{5.1} periodic
with respect to~$x$ and write these solutions in an explicit form. We
prove a uniqueness theorem in the class~$S'$ of tempered distributions
(generalized functions of slow growth) (Theorem~6.1). We show that the
spectrum of the integral operator~$K$ is continuous and is concentrated
in~$[0,1]$, and compute the (generalized) eigenfunctions. For the Hermite
coefficients $a_n=(\pfi,H_n)_1$, \ $n=2,3,\dots$, of the solution~$\pfi$
we deduce an infinite triangular linear system of equations~\thetag{6.8}.
In~\S\,7 we study the boundary-value problem~\thetag{1.3},~\thetag{1.5}
with an even~$p$ and investigate the behaviour of its solution in the
neighbourhood of real zeros of the entire function~$\pfi^p(z)$, \
$z=t+iy$, and in the neighbourhood of discontinuities of the first kind of
the solution~$\pfi(t)$~(Theorem~7.1). It is still an open question whether
this problem has solutions and whether a solution can have discontinuities
of the first kind. We also study the ramification of zeros of the
interpolating function $u(1-\veps,t-t_0)$ in the~$\veps$-neighbourhood,
$\veps>0$, of the zero~$t_0$ of~$\pfi^p(t)$ (Theorem~7.2). In~\S\,8 we
consider the special case when~$p=2$. We deduce an infinite non-linear
system of equations~\thetag{8.3} in the coefficients~$a_n$ of solutions of
equation~\thetag{1.3}. We consider the 3-approximation. In~\S\,9 we deduce
a similar system of equations~\thetag{9.4} for the solution of
problem~\thetag{1.3},~\thetag{1.5} with~$p = 2$.  We consider the
3-approximation. In~\S\,10 we study the structure of solutions of the
boundary-value problem~\thetag{1.3},~\thetag{1.6} with an odd~$p$. In
particular, we study their behaviour in the neighbourhood of real zeros of
the entire function~$\pfi^p(z)$ under the assumption that this problem has
solutions (Theorem~10.1). We point out that
problem~\thetag{1.3},~\thetag{1.6}~\cite{9} has an odd continuous
solution.

If $\pfi(t)$ is a solution of equation~\thetag{1.3}, then $\pfi(t+t_0)$
also is a solution (for every~$t_0$). Therefore, this equation cannot have
precisely one solution~$\pfi(t)$. The question whether the shifts
of~$\pfi(t)$ by~$t$ (that is, the~$\pfi(t+t_0)$, \ $t_0\in\RR$) exhaust
the set of solutions of equation~\thetag{1.3}, remains open.

It is most natural to investigate the integral equation~\thetag{1.3},
using the Hermite polynomials, since its kernel is the generating function
for these polynomials (\S\,4). This is the reason why we use these
polynomials in our construction of solutions (exact and approximate) of
equation~\thetag{1.3}. Let us note that Hermite polynomials were used
in~\cite{13}--\cite{15} in the theory of~$D$-brane perturbations as well
as in the study of more complicated problems concerning the tachyons for
both open and closed strings.

We shall need the scale of weighted separable Hilbert spaces~$L_2^\alpha$,
\ $0<\alpha<\infty$, consisting of functions on~$\RR$ measurable and
square integrable with respect to the measure
$$
d\mu_\alpha(t)=\sqrt{\frac\alpha\pi}\,e^{-\alpha t^2}\,dt, \qquad
\INT d\mu_\alpha(t)=1, \qquad
\alpha>0,
$$
with the following scalar product and norm:
$$
(f,g)_\alpha=\INT f(t)\myl{g}(t)\,d\mu_\alpha(t), \qquad
\|f\|_\alpha=\sqrt{(f,f)_\alpha}\,, \qquad
f,g\in L_2^\alpha,  \quad
\alpha>0.
$$

The embedding $L_2^\alpha\mys L_2^\beta$, \ $\alpha<\beta$, is dense and
continuous, and
$$
\|f\|_\beta\le\|f\|_\alpha, \qquad
f\in L_2^\alpha.
\tag1.7
$$

The following assertion will be used in the study of
problems~\thetag{1.3}, \thetag{1.5} and~\thetag{1.3}, \thetag{1.6}.

\proclaim{Assertion 1.1} If\/ $\pfi$~is a solution of equation~\thetag{1.3}
such that
$$
\lim_{t\to\infty}\pfi(t)=a,
\qquad
|a|<\infty,
$$
then\/ $a\!=\!0$ or\/ $a\!=\!1$ if\/~$p$ is even 
and\/~$a\!=\!0$ or\/ $a\!=\!\pm1$ 
if\/~$p$ is odd,  $\lim_{t\to\infty}(\pfi^p)'(t)=\nomathbreak0$. 
If\/~$a\ne0$, then\/ $\lim_{t\to\infty}\pfi'(t)=0$.
\endproclaim

\demo{Proof} We deduce from equation~\thetag{1.3} the following chain of
equalities:
$$
\alignat1
\lim_{t\to\infty}\pfi^p(t)
&=\biggl[\lim_{t\to\infty}\pfi(t)\biggr]^p=a^p
=\lim_{t\to\infty}\frac1{\sqrt{\pi}}\INT\pfi(\tau)e^{-(t-\tau)^2}\,d\tau
\\
&=\lim_{t\to\infty}\frac1{\sqrt{\pi}}\INT\pfi(t-u)e^{-u^2}\,du
=\frac1{\sqrt{\pi}}\INT\lim_{t\to\infty}\pfi(t-u)e^{-u^2}\,du=a,
\endalignat
$$
whence $a=0$ or $a=1$ if~$p$ is even and~$a=0,\pm1$ if~$p$ is odd.
Further, we have
$$
\alignat1
&\lim_{t\to\infty}(\pfi^p)'(t)=-2\lim_{t\to\infty}\frac1{\sqrt{\pi}}
\INT\pfi(\tau)(t-\tau)e^{-(t-\tau)^2}\,d\tau
\\
&\qquad
=-2\lim_{t\to\infty}\frac1{\sqrt{\pi}}\INT\pfi(t-u)ue^{-u^2}\,du
=\frac{-2}{\sqrt{\pi}}\INT\lim_{t\to\infty}\pfi(t-u)ue^{-u^2}\,du
\\
&\qquad
=-\frac2{\sqrt{\pi}}\,a\INT ue^{-u^2}\,du=-\frac2{\sqrt{\pi}}\,a\cdot0=0.
\endalignat
$$
If $a\ne0$, then $\lim_{t\to\infty}\pfi'(t)=0$, since
$$
\lim_{t\to\infty}(\pfi^p)'(t)=p\lim_{t\to\infty}\pfi^{p-1}(t)\pfi'(t)
=pa^{p-1}\lim_{t\to\infty}\pfi'(t)=0.
$$
We passed to the limit under the integral sign, using Lebesgue's theorem
and estimate~\thetag{1.4}.
\enddemo

We shall write $a\equiv b$ if the integers $a$ and $b$ are both even or
both odd, and \ $a\not\equiv b$ if one of them is even and the other is
odd.

\head \S\,2. Hermite polynomials
\endhead

{\it Hermite polynomials\/} are defined to be the polynomials~\cite{10}
$$
H_n(x)=(-1)^ne^{x^2}\frac{d^n}{dx^n}\,e^{-x^2},  \qquad
n=0,1,\dots,
\tag2.1
$$
whence $H_0(x)=1$, \ $H_1(x)=2x$, \ $H_2(x)=4x^2-2$, \
$H_3(x)=8x^3-12x,\dots$\,. They form a complete orthogonal system in the
Hilbert space~$L_2^1$, and
$$
\|H_n\|_1^2=\INT H_n^2(x)\,d\mu_1(x)=2^n n!.
\tag2.2
$$
Any $f\in L_2^1$ can be expanded in Hermite polynomials:
$$
f(x)=\sum_{n=0}^\infty(f,H_n)_1\frac{H_n(x)}{2^nn!} \quad \text{in} \quad
L_2^1, \tag2.3
$$
and the Parseval--Steklov equality holds:
$$
\|f\|_1^2=\sum_{n=0}^\infty|(f,H_n)_1|^2\frac1{2^nn!}\,.
\tag2.4
$$

The following equalities hold:
$$
(x^m,H_n)_1=\cases 2^{n-m}m!\,\biggl(\dfrac{m-n}2\,!\biggr)^{-1}, &m\ge n
\text{\quad and\quad} n\equiv m,
\\
0, &m<n \text{\quad or\quad} n\not\equiv m.
\endcases
\tag2.5
$$
Here we used formula 2.20.3.4 in~\cite{11} (p.~487 of the Russian version)
with $\alpha=m+1$, \ $p=c=\nomathbreak 1$.

The values of the polynomials at~$x=0$ are given by the formulae
$$
H_{2n}(0)=(-1)^n\frac{(2n)!}{n!}\,, \qquad
H_{2n+1}(0)=0, \qquad
n=0,1,\dots\,.
\tag2.6
$$

The expansion in powers of~$x$ (see~\cite{12}, formula 8.950, p.~1047 of
the Russian version) has the form
$$
H_n(x)=n!\sum\Sb m=0 \\ m\equiv n\endSb^n
c_{n,m}x^m, \qquad
n=0,1,\dots,
\tag2.7
$$
where
$$
c_{n,m}=\cases
(-1)^{\frac{n-m}2}\dfrac{2^m}{m!}\biggl(\dfrac{n-m}2\,!\biggr)^{-1}, &
m\equiv n\text{\quad and\quad} m\le n,
\\
0, &m\not\equiv n \text{\quad or\quad} m>n.
\endcases
\tag2.8
$$
In particular, we have
$$
\gathered
c_{n,n}=\frac{2^n}{n!}\,, \qquad
c_{2n,0}=\frac{(-1)^n}{n!}\,, \qquad
c_{2n+1,1}=2\frac{(-1)^n}{n!}\,,
\\
c_{2n,2}=-2\frac{(-1)^n}{(n-1)!}\,, \qquad
c_{2n+1,3}=-\frac{4(-1)^n}{3(n-1)!}\,.
\endgathered
\tag2.9
$$

The following asymptotic formulae hold:
$$
H_n(x)=(2x)^n\bigl[1+O\bigl(x^{-2}\bigr)\bigr], \qquad
x\to\pm\infty.
\tag2.10
$$

The {\it modified Hermite polynomials\/}~are defined to be the polynomials
$$
V_n(x)=2^{-n/2} H_n\biggl(\frac{x}{\sqrt2}\biggr), \qquad n=0,1,\dots\,.
\tag2.11
$$
In particular, $V_0(x)=1$, \ $V_1(x)=x$, \ $V_2(x)=x^2-1$, \
$V_3(x)=x^3-3x,\dots$\,.

These polynomials form a complete orthogonal system in the Hilbert
space~$L_2^{1/2}$, and
$$
\|V_n\|_{1/2}^2=\INT V_n^2(x)\,d\mu_{1/2}(x)=n!.
\tag2.12
$$

Any $f\in L_2^{1/2}$ can be expanded in modified Hermite polynomials:
$$
f(x)=\sum_{n=0}^\infty(f,V_n)_{1/2}\frac{V_n(x)}{n!} \quad \text{in} \quad
L_2^{1/2}, \tag2.13
$$
and the Parseval--Steklov equality holds:
$$
\|f\|_{1/2}^2=\sum_{n=0}^\infty|(f,V_n)_{1/2}|^2\frac1{n!}\,.
\tag2.14
$$

We have the following equalities:
$$
\alignat1 \hskip-24pt(H_m,V_n)_{1/2} &=\cases
2^nm!\biggl(\dfrac{m-n}{2}\,!\biggr)^{-1}, & m\ge n \text{\quad and\quad}
n\equiv m,
\\
0, & m<n \text{\quad or\quad} n\not\equiv m,
\endcases
\tag2.15
\\
(H_n,V_m)_1 &=\cases
(-1)^{\frac{m-n}{2}}2^{n-m}m!\biggl(\dfrac{m-n}{2}\,!\biggr)^{-1}\!\!, & m\ge
n \text{\quad\! and\quad\!} n\equiv m,
\\
0, & m<n \text{\quad or\quad} n\not\equiv m.
\endcases
\tag2.16
\endalignat
$$
We used here formula 2.20.16.4 in~\cite{11} (p.~502 of the Russian version)
with $p=1/2$, \ $b=1$, \ $c=\frac1{\sqrt2}$ and $p=c=1$, \
$b=\frac1{\sqrt2}$\,.

We can express $V_n$ in terms of~$H_n$ and vice versa, using
formulae~\thetag{2.3}, \thetag{2.13}, \thetag{2.15} and~\thetag{2.16}:
$$
\gather V_n(x)=2^{-n}n!\sum_{\Sb m=0\\m\equiv
n\endSb}^n(-1)^{\frac{n-m}{2}}
\biggl(\frac{n-m}{2}\,!\biggr)^{-1}\frac{H_m(x)}{m!}\,, \tag2.17
\\
H_n(x)=n!\sum_{\Sb m=0\\m\equiv n\endSb}^n\frac{2^m}{m!}
\biggl(\frac{n-m}{2}\,!\biggr)^{-1}V_m(x). \tag2.18
\endgather
$$

The integral representation for the modified Hermite polynomials has the
form:
$$
V_n(x)=\sqrt{\frac2\pi}\INT H_n(\tau)e^{-2(t/2-\tau)^2}\,d\tau, \qquad
n=0,1,\dots\,.
\tag2.19
$$

Equality~\thetag{2.19} follows from~formula 2.20.3.16 in~\cite{11} (p.~488
of the Russian version) with $p=2$, \ $c=1$, \ $y=\tau/2$ and
formula~\thetag{2.11} defining~$V_n$.

Let $f\in L_2^{1/2}$. It follows from~\thetag{1.2} that $f\in L_2^1$,
$$
\sum_{n=0}^\infty a_n\frac{H_n(x)}{2^nn!}=f(x) =\sum_{n=0}^\infty
b_n\frac{V_n(x)}{n!}  \quad \text{in} \quad L_2^1, \tag2.20
$$
and $a_n=(f,H_n)_1$ can be expressed in terms of~$b_n=(f,V_n)_{1/2}$ and
vice versa by the formulae
$$
\alignat2
a_n
&=\sum\Sb m\ge n \\ m\equiv n \endSb^\infty
(-1)^{\frac{m-n}{2}}2^{n-m}\biggl(\frac{m-n}{2}\,!\biggr)^{-1}b_m,
&\qquad
&n=0,1,\dots,
\tag2.21
\\
b_n
&=\sum\Sb m\ge n \\ m\equiv n \endSb^\infty
2^{n-m}\biggl(\frac{m-n}{2}!\biggr)^{-1}a_m,
&\qquad
&n=0,1,\dots\,.
\tag2.22
\endalignat
$$
These formulae follow from~\thetag{2.19},~\thetag{2.15} and~\thetag{2.16}.

\head \S\,3. Properties of the operator~$K$
\endhead

We denote by~$K$ the linear integral operator in equation~\thetag{1.3}:
$$
\pfi\to(K\pfi)(t)
\equiv\frac1{\sqrt\pi}\INT e^{-(t-\tau)^2}\pfi(\tau)\,d\tau.
$$

\proclaim{Lemma 3.1} The operator\/~$K$ assigns to every function\/ $f(t)$
satisfying condition~\thetag{1.4} an entire function\/ $(Kf)(z)$ with the
estimate
$$
|(Kf)(z)|\le\frac{C}{\sqrt\veps}\,\exp\biggl\{y^2
+\biggl(\frac1\veps-1\biggr)t^2\biggr\},  \qquad
z=t+iy.
\tag3.1
$$
\endproclaim

\demo\nofrills{Proof\/} \ follows immediately from~\thetag{1.4}:
$$
\align
|(Kf)(z)|
&\le\frac{C}{\sqrt\pi}\INT e^{(1-\veps)\tau^2}|e^{-(t-\tau)^2}|\,d\tau
\\
&=\frac{C}{\sqrt\pi}\,e^{y^2-t^2}\INT e^{-\veps\tau^2+2t\tau}\,d\tau
=\frac{C}{\sqrt\veps}\,e^{y^2+(1/\veps-1)t^2}.
\endalign
$$
\enddemo

\proclaim{Lemma 3.2} The operator\/~$K$ assigns to\/~$f\in L_2^\alpha$, \
$0<\alpha<2$, an entire function\/ $(Kf)(z)$ with the estimate
$$
|(Kf)(z)|\le\|f\|_\alpha(2-\alpha)^{-1/4}
\exp\biggl\{y^2+\frac\alpha{2-\alpha}t^2\biggr\}, \qquad
z=t+iy.
\tag3.2
$$
\endproclaim

\demo\nofrills{Proof\/} \ follows from the Cauchy--Bunyakovskii inequality
(applied to~$(Kf)(z)$) and the following estimates:
$$
\alignat1
|(Kf)(z)|
&\le\frac1{\sqrt\pi}\INT|f(\tau)|e^{-\alpha\tau^2/2}
\biggl|\exp\biggl\{-z^2+2z\tau-
\biggl(1-\frac\alpha2\biggr)\tau^2\biggr\}\biggr|\,d\tau
\\
&\le\|f\|_\alpha\biggl[\frac1{\sqrt\pi}\INT
|\exp\{2z^2+4t\tau-(2-\alpha)\tau^2\}|\,d\tau\biggr]^{1/2}
\\
&\le\|f\|_\alpha\biggl[\frac1{\sqrt\pi}\INT
\exp\{2y^2-2t^2+4t\tau-(2-\alpha)\tau^2\}\,d\tau\biggr]^{1/2}
\\
&=\|f\|_\alpha\exp\biggl\{y^2+\frac\alpha{2-\alpha}t^2\biggr\}
\biggl[\frac1\pi\INT\exp\biggl\{(2-\alpha)
\biggl(\tau-\frac{2t}{2-\alpha}\biggr)^2\biggr\}d\tau\biggr]^{1/2}.
\endalignat
$$
\enddemo

\proclaim{Lemma 3.3} The operator\/~$K\: L_2^\alpha\to L_2^\beta$, \
$0<\alpha<2$, \ $\beta>\frac{2\alpha}{2-\alpha}$\,, is bounded, and
$$
\|Kf\|_\beta\le\biggl(2{\alpha}-\frac{2\alpha^2}{\beta}
-\alpha^2\biggr)^{-1/4}\|f\|_\alpha,
\qquad f\in L_2^\alpha. \tag3.3
$$
\endproclaim

\demo{Proof}  We prove the lemma by writing the following chain
of equalities and inequalities for $f\in L_2^\alpha$:
$$
\alignat1 \|Kf\|^2_\beta &=\sqrt{\frac\beta\pi}\INT e^{-\beta
t^2}|(Kf)(t)|^2\,dt
\\
&=\sqrt{\frac\beta\pi}\INT e^{-(\beta+2)t^2}\bigg|\frac{1}{\sqrt{\pi}}\INT
f(\tau)e^{-(\alpha/2)\tau^2-(1-\alpha/2)\tau^2+2t\tau}\,d\tau\bigg|^2dt
\\
&\le\frac1\pi\sqrt{\frac\beta\pi}\INT e^{-(\beta+2)t^2} \INT |f(\tau)|^2
e^{-\alpha\tau^2}\,d\tau\INT e^{-(2-\alpha)\tau^2+4t\tau}\,d\tau\,dt
\\
&=\frac1\pi\,\sqrt{\frac\beta\alpha}\,\|f\|^2_\alpha \INT
e^{-\bigl(\beta+2-\frac4{2-\alpha}\bigr)t^2}\,dt\INT
e^{-(2-\alpha)\tau^2}\,d\tau
\\
&=(2\beta-2\alpha-\alpha\beta)^{-1/2}
\sqrt{\frac\beta\alpha}\,\|f\|^2_\alpha.
\endalignat
$$
\enddemo

\proclaim{Lemma 3.4} If\/ $f\in L_2^1$, then its image\/ $(Kf)(t)$ can be
expanded in the Taylor series
$$
(Kf)(t)=\sum_{n=0}^\infty a_n\frac{t^n}{n!}\,, \qquad
a_n=(f,H_n)_1,
\tag3.4
$$
which converges uniformly on every compact set in\/~$\RR$. If\/~$f\in
L_2^{1/2}$, then
$$
(Kf)(t)=\sum_{n=0}^\infty b_n\frac{H_n(t)}{2^nn!} \quad \text{in} \quad
L_2^1, \qquad b_n=(Kf,H_n)_1, \tag3.5
$$
and
$$
(Kf,H_n)_1=(f,V_n)_{1/2}, \qquad
n=0,1,\dots\,.
\tag3.6
$$
\endproclaim

\demo{Proof} By Lemma~3.3, the function $(Kf)(t)$ is the trace of an
entire function $(Kf)(z)$ for $y=0$. Hence, it can be expanded in the
Taylor series with the coefficients
$$
\align
\frac{d^n}{dt^n}(Kf)(t)|_{t=0}
&=\frac1{\sqrt\pi}\INT f(\tau)\frac{d^n}{dt^n}\,e^{-(t-\tau)^2}\,d\tau\Big|_{t=0}
\\
&=\frac1{\sqrt\pi}\INT f(\tau)(-1)^ne^{-(t-\tau)^2}H_n(t-\tau)\,d\tau|_{t=0}
\\
&=\frac1{\sqrt\pi}\INT f(\tau)(-1)^nH_n(-\tau)e^{-t^2}\,d\tau
=(f,H_n)_1=a_n.
\endalign
$$
Here we used equality~\thetag{2.1}.

Further, if~$f\in L_2^{1/2}$, then~\thetag{3.5} holds by~\thetag{2.20},
since $Kf\in L_2^1$ by Lemma~3.3. Equalities~\thetag{3.6} can be proved as
follows:
$$
(Kf,H_n)_1=(f,K^*H_n)_{1/2}=(f,V_n)_{1/2}.
$$
Here we used formula~\thetag{2.19}, which implies that~$V_n=K^*H_n$,
where~$K^*$~is the operator adjoint to~$K$.
\enddemo

\head \S\,4. Expanding solutions in Hermite polynomials
\endhead

Let $\pfi$~be a solution of equation~\thetag{1.3} belonging to~$L_2^1$,
whence $\pfi^p=K\pfi$. Putting $a_n=(\pfi,H_n)_1$, we deduce
from~\thetag{2.3} and~\thetag{2.4} that
$$
\pfi(t)=\sum_{n=0}^\infty a_n\frac{H_n(t)}{2^nn!} \quad \text{in} \quad
L_2^1, \qquad \sum_{n=0}^\infty\frac{a_n^2}{2^nn!}=\|\pfi\|^2. \tag4.1
$$
The function $\pfi^p(t)$ is the trace of the entire function
$A(z)=(K\pfi)(z)$, for which~\thetag{3.2} holds with $\alpha=1$:
$$
|A(z)|\le\|\pfi\|_1e^{|z|^2}, \qquad z\in\CC. \tag4.2
$$
By Lemma~3.4, it can be expanded in the Taylor series~\thetag{3.4}:
$$
\pfi^p(t)=\sum_{n=0}^\infty a_n\frac{t^n}{n!}\,.
\tag4.3
$$

If $\pfi\in L_2^{1/2}$ is a solution of equation~\thetag{1.3}, then
Lemma~3.3 implies that $\pfi^p\in L_2^1$ and
$$
\|\pfi^p\|_1\le \sqrt{2} \|\pfi\|_{1/2}. \tag4.4
$$
Lemma~3.4 implies that equalities in~\thetag{3.6} (integral conservation
laws) hold:
$$ (\pfi^p,H_n)_1=b_n=(\pfi,V_n)_{1/2}, \qquad
n=0,1,\dots\,. \tag4.5
$$
Therefore, $\pfi^p(t)$ can be expanded in Hermite polynomials
$$
 \pfi^p(t) =\sum_{n=0}^\infty b_n\frac{H_n(t)}{2^nn!} \quad \text{in}
\quad L_2^1, \tag4.6
$$
$\pfi(t)$~can be expanded in modified Hermite polynomials:
$$
\pfi(t) =\sum_{n=0}^\infty b_n\frac{V_n(t)}{n!} \quad \text{in} \quad
L_2^{1/2}, \tag4.7
$$
and the corresponding Parseval--Steklov equalities hold:
$$
\sum_{n=0}^\infty\frac{b_n^2}{2^nn!}=\|\pfi^p\|_1^2, \qquad
\sum_{n=0}^\infty\frac{b_n^2}{n!}=\|\pfi^p\|_{1/2}^2.
\tag4.8
$$
Let us note that the $a_n$ and~$b_n$ are related by~\thetag{2.21}
and~\thetag{2.22}.

\head \S\,5. Connection between $\pfi(t)$ and the solutions of the heat
equation
\endhead

The integral equation~\thetag{1.3} is equivalent to the following
boundary-value problem for the heat equation:
$$
\gather
u_x=\frac14\,u_{tt}, \qquad
0<x\le1, \quad
t\in\RR,
\tag5.1
\\
u(0,t)=\pfi(t), \qquad
u(1,t)=\pfi^p(t),  \quad
t\in\RR.
\tag5.2
\endgather
$$

A {\it solution of problem\/}~\thetag{5.1}, \thetag{5.2} is defined to be
any measurable function $u(x,t)$~for which~\thetag{1.4} holds, where~$C$
does not depend on~$x$. We say that~$u(x,t)$ is an {\it interpolating\/}
function between~$\pfi(t)$ and~$\pfi^p(t)$.

Let us note that if there is an interpolating function, it can be
represented by Poisson's formula for equation~\thetag{5.1}:
$$
u(x,t)=\frac1{\sqrt{\pi x}}\INT\pfi(\tau)
\exp\biggl\{-\frac{(t-\tau)^2}{x}\biggr\}\,d\tau, \qquad
0<x\le1.
\tag5.3
$$
If $\pfi$ is such that
$$
|\pfi(t)|\le C\exp\{\veps t^2\} \quad \text{for any} \quad \veps>0, \quad
t\in\RR, \tag5.4
$$
then formula~\thetag{5.3} gives its analytic continuation to the domain
$x>1$, \ $t\in\RR$ and, further, its analytic continuation with respect to
$(x,t)$ to the complex domain $T^+\myt\CC$, where~$T^+$ is the right
half-plane $\Re\zeta=x>0$.

Representation~\thetag{5.3} implies that if~$\pfi(t)$~is a solution of
problem~\thetag{1.3},~\thetag{1.5} or \thetag{1.3},~\thetag{1.6}, then
$|u(x,t)|<1$, \ $0\le x$, \ $t\in\RR$, \ $u(x,t)$ satisfies the boundary
conditions~\thetag{1.5} or~\thetag{1.6}, respectively, and~$u(x,t)>0$ for
$x>1$.

\example{Example} For the solution
$$
\pfi(t)=p^{\frac1{2(p-1)}}\exp\biggl\{\frac{p-1}{p}\,t^2\biggr\}
$$
of equation~\thetag{1.3} we have the following interpolating function:
$$
u(x,t)=p^{\frac1{2(p-1)}}\biggl(1-x+\frac xp\biggr)^{-\frac12}
\exp\biggl\{\frac{t^2(p-1)}{p-xp+x}\biggr\}.
$$
\endexample

\proclaim{Theorem 5.1} Let\/ $u(x,t)$~be an interpolating function between
the solution\/~$\pfi$ and its power\/~$\pfi^p$ for problems~\thetag{1.3},
\thetag{1.5} and\/~\thetag{1.3},~\thetag{1.6}. Then
$$
\gather
\INT\pfi^2(t)[1-\pfi^{2p-2}(t)]\,dt
=\frac12\int_0^1\INT u_t^2(x,t)\,dt\,dx,
\tag5.5
\\
\INT[u(x,t)-\pfi(t)]\,dt=0,
\quad\;
\INT[u(x+1,t)-\pfi^p(t)]\,dt=0,
\quad\;
x\ge0
\tag5.6
\endgather
$$
{\rm(}a conservation law{\rm)}.
\endproclaim

\goodbreak

\demo{Proof} Multiplying equation~\thetag{5.1} by~$u(x,t)$, integrating
both sides over the rectangle $0\le x\le1$, \ $a\le t\le b$, and taking
into account~\thetag{5.2}, we obtain the following chain of equalities:
$$
\alignat1
&\int_0^1\int_a^buu_x\,dt\,dx=\frac12\int_a^b[u^2(1,t)-u^2(0,t)]\,dt
\\
&\qquad
=\frac12\int_a^b\pfi^2(t)[\pfi^{2p-2}(t)-1]\,dt
=\frac14\int_0^1\int_a^buu_{tt}\,dt\,dx
\\
&\qquad
=-\frac14\int_0^1\int_a^bu_t^2\,dt\,dx
+\frac14\int_0^1[u(x,b)u_t(x,b)-u(x,a)u_t(x,a)]\,dx.
\tag5.7
\endalignat
$$
Using~\thetag{5.3},~\thetag{1.5} and~\thetag{1.6}, we obtain that
$$
\gather
{\align
\lim_{b\to\infty}u(x,b)
&=\frac1{\sqrt\pi}\INT\lim_{b\to\infty}\pfi(b-\sqrt{x}\,v)e^{-v^2}\,dv=1,
\\
\lim_{b\to\infty}u_t(x,b)
&=-\frac2{\sqrt{x\pi}}\INT\lim_{b\to\infty}\pfi(b-\sqrt{x}\,v)ve^{-v^2}\,dv=0,
\endalign}
\\
|u_t(x,b)|<\frac2{\sqrt{x\pi}}\,, \qquad
0<x\le1.
\endgather
$$
(The passage to the limit under the integral sign is possible by
Lebesgue's theorem.) Similar relations hold for~$u(x,a)$ and~$u_t(x,a)$.
Taking into account these relations and passing to the limit
in~\thetag{5.7}, we obtain~\thetag{5.5}.

Equality~\thetag{5.6} can be proved as follows. Integrating
equation~\thetag{5.1} over the rectangle $0\le x\le X$, \ $a\le t\le b$,
we obtain, as before, the chain of equalities
$$
\alignat1
&\int_a^b[u(X,t)-\pfi(t)]\,dt=\frac14\int_0^X[u_t(x,b)-u_t(x,a)]\,dx
\\
&\qquad
=-\frac1{2\sqrt\pi}\int_0^X\frac1{x\sqrt x}\INT\Bigl[(b-\tau)e^{-\frac{(b-\tau)^2}x}
-(a-\tau)e^{-\frac{(a-\tau)^2}x}\Bigr]\pfi(\tau)\,d\tau\,dx
\\
&\qquad
=-\frac1{2\sqrt\pi}\int_0^X\frac1{\sqrt x}\INT
ve^{-v^2}[\pfi(b-v\sqrt x\,)-\pfi(a-v\sqrt x\,)]\,dv\,dx.
\endalignat
$$
Passing here to the limit as~$a\to-\infty$ and~$b\to\infty$ and taking
into account the boundary conditions~\thetag{1.5} and~\thetag{1.6}, we
obtain~\thetag{5.6}.
\enddemo

\proclaim{Corollary 5.1} {\rm(i)} The integrals
$$
\alignedat2 &\int_{-\infty}^0\pfi^2(t)\,dt, \quad
\int_0^\infty\bigl(1-\pfi^{2p-2}(t)\bigr)\,dt &\quad
&\text{for\quad\thetag{1.3}, \thetag{1.5}},
\\
&\INT\bigl(1-\pfi^{2p-2}(t)\bigr)\,dt & \quad &\text{for\quad\thetag{1.3},
\thetag{1.6}}
\endalignedat
\tag5.8
$$
converge if and only if
$$
\int_0^1\INT u_t^2(x,t)\,dt\,dx<\infty.
$$

{\rm(ii)} Inequality~\thetag{5.6} with\/ $x=1$ implies that
$$
\INT[\pfi(t)-\pfi^p(t)]\,dt=0.
\tag5.9
$$

{\rm(iii)} The following integrals converge for
problem~\thetag{1.3},~\thetag{1.5}:
$$
\int_0^\infty[1-\pfi^{p-1}(t)]\,dt, \qquad \int_0^\infty[1-\pfi(t)]\,dt,
\qquad \int_0^\infty[1-u(x,t)]\,dt, \qquad x\ge1.
$$
If\/ $\pfi(t)$ has constant sign for\/ $t<c$, then the following integrals
also converge:
$$
\int_{-\infty}^0\pfi(t)\,dt, \qquad
\int_{-\infty}^0u(x,t)\,dt, \qquad
x\ge1.
$$

{\rm(iv)} The following integrals converge for
problem~\thetag{1.3},~\thetag{1.6}:
$$
\gather
\INT\bigl[1-\pfi^{p-1}(t)\bigr]\,dt, \qquad
\int_0^\infty[1-\pfi(t)]\,dt, \qquad
\int_{-\infty}^0[1+\pfi(t)]\,dt,
\\
\int_0^\infty[1-u(x,t)]\,dt, \quad
\int_{-\infty}^0 [1+u(x,t)]\,dt, \qquad
x\ge0.
\endgather
$$
\endproclaim

\head \S\,6. The linear case ($p=1$)
\endhead

Consider the linear equation~\thetag{1.3}:
$$
\pfi(t)=\frac1{\sqrt\pi}\INT\pfi(\tau)e^{-(t-\tau)^2}\,d\tau.
\tag6.1
$$
It is natural to consider a more general problem for equation~\thetag{6.1}
-- the spectral problem of finding generalized eigenfunctions for the
operator~$K$ (see~\S\,3):
$$
\lambda\pfi=K\pfi, \qquad
\pfi\in L_2^\alpha, \quad
0<\alpha<2.
\tag6.2
$$

\proclaim{Theorem 6.1} The spectrum of\/~$K$ is continuous and concentrated
in\/~$[0,1]$, and to every eigenvalue\/ $\lambda_\xi=e^{-\xi^2/4}$, \
$-\infty<\xi<\infty$, there correspond two eigenfunctions:
$$
\pfi_\xi(t)=\cases \cos(\xi t), \; \sin(\xi t) &\text{if\quad $\xi\ne0$},
\\
1, \; t &\text{if\quad $\xi=0$}.
\endcases
\tag6.3
$$

These functions exhaust the solutions of equation~\thetag{6.2} with
$\lambda=\lambda_\xi$ in the class~$S'$ of tempered distributions. The set
of these functions is dense in~$L_2^\alpha$, \ $\alpha>0$.
\endproclaim

\demo{Proof} Let $\pfi\in S'$~be a solution of equation~\thetag{6.2}
written as~$\lambda\pfi=\pfi\star\,d\mu_1$. Passing to the Fourier
transform~$\myw\pfi(\xi)$ and using the theorem on the Fourier transform
of a convolution~\cite{13}, we obtain the equation
$$
\lambda\myw\pfi(\xi)=e^{-\xi^2/4}\myw\pfi(\xi),
\tag6.4
$$
whence we deduce that either $\myw\pfi(\xi)=0$ or for
$\lambda=\lambda_{\xi_0}$ the support of~$\myw\pfi$ consists of the
points~$\pm\xi_0$ (if $\xi_0=0$, then this support consists of the single
point~$0$). In this case~$\myw\pfi$ can be represented as
$$
\myw\pfi(\xi)=\sum_{k=0}^nc_k\delta^{(k)}(\xi-\xi_0)
+d_k\delta^{(k)}(\xi+\xi_0)
\tag6.5
$$
with some (complex) $c_k$ and~$d_k$. Substituting~\thetag{6.5}
into~\thetag{6.2} and using the fact that the distributions (generalized
functions)~$\delta^{(k)}$, \ $k=0,1,\dots,n$, are linearly independent, we
obtain, after some natural transformations, that $c_k=d_k=0$, \
$k=1,2,\dots,n$, if~$\xi_0\ne0$ and~$c_k=d_k=0$, \ $k=2,3,\dots$, \
$d_1=0$ if~$\xi_0=0$. Therefore,
$$
\myw\pfi(\xi)=\cases c_0\delta(\xi-\xi_0)+d_0\delta'(\xi-\xi_0) &\text{if}
\quad \xi_0\ne0,
\\
c_0\delta(\xi)+c_1\delta'(\xi) &\text{if} \quad \xi_0=0,
\endcases
$$
which implies that~\thetag{6.3} holds.

The set of solutions~\thetag{6.3} is dense in~$L_2^\alpha$, \ $\alpha>0$.
Indeed, if there is an $f\in L_2^\alpha$ orthogonal to the functions
defined by formula~\thetag{6.3}, that is,
$$
(f,\pfi_\xi)_\alpha=\frac1{\sqrt\pi}\INT f(\tau)\,
e^{-\alpha\tau^2}e^{i\xi\tau}\,d\tau=0
$$
for all $\xi\in\RR$, then we have $f(\tau)e^{-\alpha\tau^2}=0$, since
$f(\tau)e^{-\alpha\tau^2}\!\in L_1(\RR)$. Hence, $f(\tau)=0$ almost
everywhere in~$\RR$.
\enddemo

\proclaim{Corollary 6.1} The operator\/~$K$ is bounded, self-adjoint and
positive in\/~$L_2(\RR)$, and
$$
\|Kf\|\le\|f\|, \qquad
(Kf,f)\ge0, \qquad
f\in l_2(\RR).
$$
\endproclaim

By Theorem 6.1, equation~\thetag{6.1} has two linearly independent
solutions in~$S'$: $1$ and~$t$. The solutions of equation~\thetag{6.1}
belonging to~$L_2^{1/2}$, if there are any, can be expanded in
series~\thetag{4.1}, \thetag{4.3}, \thetag{4.6} and~\thetag{4.8}
with~$a_n=b_n=(\pfi,H_n)_1$, \ $n=0,1,\dots$\,. The infinite
systems~\thetag{2.21} and~\thetag{2.22} take the form
$$
\alignat2
\sum\Sb m=n+2 \\ m\equiv n \endSb^\infty
(-1)^{\frac{m-n}{2}}2^{n-m}\biggl(\frac{m-n}{2}\,!\biggr)^{-1}a_m
&=0,
&\qquad
&n=0,1,\dots,
\tag6.6
\\
\sum\Sb m=n+2 \\ m\equiv n \endSb^\infty
2^{n-m}\biggl(\frac{m-n}{2}\,!\biggr)^{-1}a_m
&=0,
&\qquad
&n=0,1,\dots\,.
\tag6.7
\endalignat
$$

The linear systems~\thetag{6.6} and~\thetag{6.7} are triangular, they do
not contain~$a_0$ and~$a_1$ and are decomposed into four independent
block-triangular systems corresponding to the values of the residue of~$n$
modulo~4.

Indeed, adding systems~\thetag{6.6} and~\thetag{6.7} together and
subtracting one of them from the other, we obtain four groups (for
$\myk=0,1,2,3$) of independent block-triangular systems in~$a_m$, \
$m=2+4k+4l+\myk$, \ $l,k=0,1,\dots$:
$$
\sum_{l=0}^\infty a_{2+4k+4l+\myk}\frac1{2^{4l}(2l+1)!}=0, \qquad
\sum_{l=0}^\infty a_{2+4k+4l+\myk}\frac1{2^{4l}(2l+2)!}=0.
\tag6.8
$$

It turns out that there are non-trivial solutions of system~\thetag{6.8}
(with the exception of the case when~$a_0$ and~$a_1$ are arbitrary
and~$a_2=a_3=\dots=0$) such that $\sum a_m^2/m!<\infty$, that is,
equation~\thetag{6.1} has the following solutions belonging
to~$L_2^{1/2}$:
$$
\pfi_k^\pm(t)=e^{\pm2\sqrt{k\pi}\,t}\cos\bigl(2\sqrt{k\pi}\,t\bigr), \qquad
k=0,1,\dots\,.
\tag6.9
$$
This follows from the fact that the heat equation~\thetag{5.1} has the
following solutions periodic with respect to~$x$ with period~1:
$$
u_k^\pm(x,t)=e^{\pm2\sqrt{k\pi}\,t}\cos\bigl(2\sqrt{k\pi}\,t\pm2k\pi
x\bigr), \qquad k=0,1,\dots\,. \tag6.10
$$
We have $\pfi_k^\pm(t)=u_k^\pm(1,t)$.

\head \S\,7. The case when $p$ is even ($p=2q$)
\endhead

Equation~\thetag{1.3} takes the form
$$
\pfi^{2q}(t)=\frac1{\sqrt\pi}\INT\pfi(\tau)e^{-(t-\tau)^2}\,d\tau, \qquad
q=1,2,\dots\,.
\tag7.1
$$
If $\pfi(t)$ is a solution of equation~\thetag{7.1}, then $\pfi(-t)$
and~$\pfi(t+t_0)$ also are solutions of this equation (for all~$t_0$).

Assume that problem~\thetag{7.1}, \thetag{1.5} has a piecewise continuous
solution~$\pfi(t)$. By Theorem~2 in~\cite{9}, we have \ $|\pfi(t)|<1$,
and~$\pfi(t)$ satisfies the equation
$$
\pfi^{2q}(t)=A(t),
\tag7.2
$$
where $A(t)\ge0$~is the trace of an entire function $A(z)$~for which
estimate~\thetag{4.2} holds. Equation~\thetag{7.2} has two real solutions
in the neighbourhood of every~$t$:
$$
\pfi(t)=\pm A^{\frac1{2q}}(t).
\tag7.3
$$
Therefore, the global structure of the solution~$\pfi(t)$ depends on its
points~$T_k$ of discontinuity of the first kind and on the real
zeros~$t_k$ of the entire function~$A(z)$. The sets~$\{T_k\}$
and~$\{t_k\}$ are bounded above and at most countable.

By~\thetag{7.2}, the function $\pfi^{2q}(t)$ has the following
representation in the neighbourhood of~$t_k$:
$$
\pfi^{2q}(t)
=\frac{a_{2\sigma_k}}{(2\sigma_k)!}(t-t_k)^{2\sigma_k}[1+O(|t-t_k|)],
\tag7.4
$$
where $a_{2\sigma_k}>0$ and the multiplicity of the zero~$2\sigma_k$ is an
even number. Hence, $\pfi(t)$~can be represented in the neighbourhood
of~$t_k$ as follows:
$$
\pfi(t)=\pm\biggl[\frac{a_{2\sigma_k}}{(2\sigma_k)!}\biggr]^{\frac1{2q}}
|t-t_k|^{\frac{\sigma_k}q}[1+O(|t-t_k|)], \tag7.5
$$
and
$$
\frac{2^{2\sigma_k}}{\sqrt\pi}\INT\pfi(\tau)(t_k-\tau)^n
e^{-(t_k-\tau)^2}\,d\tau=\cases
a_{2\sigma_k}>0,
& n=2\sigma_k,
\\
0,
& n=0,1,\dots,2\sigma_k-1.
\endcases
\tag7.6
$$

By~\thetag{7.3}, the function $\pfi(t)$ can be represented in the
neighbourhood of the point~$T_k$ of discontinuity of the first kind as
follows:
$$
\pfi(t)=\pm\sgn t\,A^{\frac1{2q}}(t). \tag7.7
$$
Therefore, it has the saltus
$$ 2A^{\frac1{2q}}(T_k+0) \quad \text{or}
\quad -2A^{\frac1{2q}}(T_k+0)
$$
at $T_k$. Consider the different cases.

(a) The function $A(t)$ has no zeros, whence $A(t)>0$ for all~$t$. By
Theorem~3 in~\cite{9}, \ $\pfi(t)=A^{1/(2q)}(t)>0$ cannot be a solution of
problem~\thetag{7.1}, \thetag{1.5}. Therefore, its sign cannot be
constant. Hence, there is a~$T_0$ in whose neighbourhood~$\pfi(t)$ has
representation~\thetag{7.7}.

If $\pfi(t)$ has only one point of discontinuity of the first kind
(at~$T_0$), that is, $\pfi(t)>0$ for $t>T_0$ and~$\pfi(t)<0$ for $t<T_0$,
then
$$
A'(T_0)=-\frac2{\sqrt\pi}\INT\pfi(\tau)(T_0-\tau)e^{-(T_0-\tau)^2}\,d\tau>0.
\tag7.8
$$

(b) Assume that $A(t)$ has only one zero $t_0$. Hence,
$A(t_0)=A'(t_0)=\nomathbreak 0$. Then problem~\thetag{7.1}, \thetag{1.5}
has no continuous solution~$\pfi(t)$. Indeed, in this case $\pfi(t)>0$ for
$t>t_0$ and~$\pfi(t)<0$ for $t<t_0$. Formula~\thetag{7.8} with~$t_0$
instead of~$T_0$ implies that $A'(t_0)>0$, which contradicts the relation
$A'(t_0)=0$. Hence, points of discontinuity of the first kind can exist in
this case as well.

(c) Assume that $A(t)$ has only two zeros: $t_0=0$ and~$t_1<0$. Then
$A(0)=A'(0)=0$ and~$A(t_1)=A'(t_1)=0$. A priori the following three cases
are possible for continuous solutions:
$$
\pfi(t)=\cases
A^{\frac1{2q}}(t),
& t>0,
\\
\pm A^{\frac1{2q}}(t),
& t_1<t<0,
\\
\pm A^{\frac1{2q}}(t),
& t<t_1.
\endcases
\tag7.9
$$

We claim that in reality of all these four cases only the following case
can occur:
$$
\pfi(t)=\cases
A^{\frac1{2q}}(t),
& t>0,
\\
-A^{\frac1{2q}}(t),
& t_1<t<0,
\\
A^{\frac1{2q}}(t),
& t<t_1.
\endcases
\tag7.10
$$

Indeed, the case $(+,+)$ contradicts Theorem~3 in~\cite{9}. The
case~$(-,-)$ contradicts the relation $A'(0)=0$ (see~\thetag{7.8} with
$T_0=0$), and the case $(+,-)$~contradicts the relation $A'(t_1)=0$
(see~\thetag{7.8} with $T_0=t_1$).

(d) Assume that the solution $\pfi(t)$ is continuous and the
function~$\pfi^{2q}(t)$ has a zero~$t_0$ of multiplicity~$2n$.
Then~$\pfi(t)$ has at least~$2n$ sign changes.

Indeed, we can assume without loss of generality that $t_0=0$.
By~\thetag{7.6}, we have
$$
\INT\pfi(\tau)\tau^ke^{-\tau^2}\,d\tau=0, \qquad
k=0,1,\dots,2n-1.
\tag7.11
$$
Assume the contrary: let $\pfi(t)$ have $m<2n$ sign changes. Let
$(a_k,b_k)$, \ $k=1,2,\dots,m\le2n-1$, be the intervals on which~$\pfi(t)$
is negative (on the complementary segments it is non-negative). There is a
polynomial of degree~$m$ such that $P(t)>0$ if~$t\in(a_k,b_k)$
and~$P(t)\ge0$ on the complementary segments. We have
$$
\INT\pfi(\tau)P(\tau)e^{-\tau^2}\,d\tau>0,
$$
which contradicts~\thetag{7.11}.

\goodbreak
The results obtained above can be stated as follows.

\proclaim{Theorem 7.1} If\/ $\pfi$~is a solution of problem~\thetag{7.1},
\thetag{1.5}, then

{\rm(i)} \ $\pfi(t)$ can have discontinuity of the first kind at\/~$T_k$
with saltuses\/ $\pm2\pfi(T_k+0)$ if\/~$\pfi^{2q}(t)$ has at most one zero,

{\rm(ii)} \  $\pfi(t)$~has structure~\thetag{7.10} if it is continuous
and\/~$\pfi^{2q}(t)$ has only two zeros\/ $t_0$ and\/~$t_1$,

{\rm(iii)} in the neighbourhood of\/~$t_k$ \ $\pfi(t)$ has
representation~\thetag{7.5}, and\/~\thetag{7.6} holds,

\rom{(iv)} if\/~$\pfi(t)$ is continuous, then the number of its sign changes
does not exceed the number of zeros of\/~$\pfi^{2q}(t)$ and is greater than
or equal to\/~$\sup_k2\sigma_k$, where\/~$2\sigma_k$~is the multiplicity of
the zero\/~$t_k$,

\rom{(v)} \ $\pfi(t)$~is piecewise real-analytic everywhere, with the
exception of its zeros and the points of discontinuity of the first kind.
\endproclaim

The following theorem holds for the zeros of the interpolating
function~$u(x,t)$.

\proclaim\nofrills{Theorem 7.2} \ {\rm(on the branching of zeros)}. Let\/
$u(x,t)$~be an interpolating function between the solution\/~$\pfi$ of
problem~\thetag{5.1},~\thetag{5.2} and\/~$\pfi^{2q}$ such that\/ $u(1,t)$ has
a zero of multiplicity\/~$2n$ at\/~$t=0$. Then the equation
$$
u(1-\veps,t)=0 \quad \text{as} \quad \veps\to+0 \tag7.12
$$
has precisely $2n$ simple real roots
$$
t_k^\pm(\veps)=\frac12\,\lambda_k^\pm\sqrt\veps+O(\veps), \qquad
k=1,2,\dots,n,
\tag7.13
$$
where\/ $\lambda_k^\pm$ are the roots of the equation
$$
\sum_{m=0}^n(-1)^m\frac{\lambda^{2n-2m}}{(2n-2m)!\,m!}=0. \tag7.14
$$
\endproclaim

 \demo{Proof} \  The results of \S\,5 imply that $u(x,t)$ has a
holomorphic continuation to~$T^+\myt\nomathbreak\CC$. 
Since $u(1,t)=\pfi^{2q}(t)>0$ is a positive function, the multiplicity 
of its zero $t=0$ is even ($2n$), whence
$$
u(1,0)=u'(1,0)=\dots=u^{(2n-1)}(1,0)=0, \qquad
u^{(2n)}(1,0)=a>0.
\tag7.15
$$

We expand $u(1-\veps,t)$ in the Taylor series about $(1,0)$:
$$
u(1-\veps,t)=\sum_{m=0}^{2n}\frac1{m!}\sum_{s=0}^m
C_m^s\frac{\myp^mu(1,0)}{\myp x^s\myp t^{m-s}}(-\veps)^st^{m-s}+R_1(\veps,t),
\tag7.16
$$
where the residual term $R_1\in C^\infty$ can be computed by the formula

$$
R_1=\frac1{(2n+1)!}\sum_{s=0}^{2n+1}C_{2n+1}^s
\frac{\myp^{2n+1}u(1-\theta\veps,\theta t)}{\myp x^s\myp t^{2n+1-s}}
(-\veps)^st^{2n+1-s} \tag7.17
$$
with some $\theta\in(0,1)$. Taking into account that the derivatives
of~$u(x,t)$ are bounded in the neighbourhood of~$(1,0)$, we obtain the
following estimate for~$R_1$ in~\thetag{7.17}:
$$
|R_1(\veps,t)|\le C_1\sum_{s=0}^{2n+1}|\veps|^s|t|^{2n+1-s}.
$$

Using~\thetag{7.15} and the heat equation~\thetag{5.1}, we
reduce~\thetag{7.16} to the form
$$
u(1-\veps,t)=a\sum_{m=0}^n(-1)^m\frac{t^{2n-2m}}{(2n-2m)!\,m!}
\biggl(\frac\veps4\biggr)^{n-m}+R_2(\veps,t),
\tag7.18
$$
where $R_2$ is such that
$$
|R_2(\veps,t)|\le C_2\sum_{s=1}^{2n}\veps^st^{2n-s}.
$$

Making the change of variable
$$
\lambda=\frac{2t}\veps\,, \qquad t=\frac{\lambda\veps}2\,, \tag7.19
$$
we transform~\thetag{7.18} into the equation
$$
\gather
u(1-\veps,t)=a4^{-n}\veps^n\sum_{m=0}^n(-1)^m
\frac{\lambda^{2n-2m}}{(2n-2m)!\,m!}+R_3(\veps,\lambda),
\tag7.20
\\
|R_3(\veps,\lambda)|\le C_3\veps^{n+1/2}.
\tag7.21
\endgather
$$

Dividing~\thetag{7.20} by~$a4^{-n}\veps^n$, we obtain the following
equation for the zeros $t(\veps)=\lambda(\veps)\sqrt{\veps}/2$ of
$u(1-\veps,t)$ (see~\thetag{7.12}):
$$
\sum_{m=0}^n(-1)^m\frac{\lambda^{2n-2m}}{(2n-2m)!\,m!}=R_4(\veps,\lambda), \qquad
|R_4(\veps,\lambda)|\le C_4\sqrt{\veps}\,.
\tag7.22
$$

The equation
$$
\sum_{m=0}^n(-1)^m\frac{\Lambda^{n-m}}{(2n-2m)!\,m!}=0
\tag7.23
$$
has $n$ positive roots $\Lambda_k$, \ $k=1,2,\dots,n$. Hence,
equation~\thetag{7.14} has~$2n$ real roots
$\lambda_k^\pm=\pm\sqrt{\Lambda_k}$\,, \ $k=1,2,\dots,n$.
Equation~\thetag{7.22} takes the following form in the neighbourhood
of~$\lambda_m^+$:
$$
\lambda-\lambda_k^+ =\frac{r(\veps,\lambda)}
{(\lambda-\lambda_k^-)\prod_{i\ne k}(\lambda^2-\Lambda_i)} \equiv
R_5(\veps,\lambda). \tag7.24
$$
By~\thetag{7.23}, the residual term $R_5$ has the following properties:
$$
|R_5(\veps,\lambda)|\le C_5\sqrt\veps\,, \qquad
|R_5(\veps,\lambda)-R_5(\veps,\lambda')|\le C_5\sqrt\veps\,|\lambda-\lambda'|.
$$
(In the neighbourhood of~$\lambda_k^-$ equation~\thetag{7.22} takes a
similar form). Using the contraction map principle, we obtain that
equation~\thetag{7.24} has precisely one continuous solution
$$
\lambda_k^+(\veps)=\lambda_k^++O(\sqrt\veps\,)
$$
in the neighbourhood of~$\lambda_k^+$. Using~\thetag{7.19}, we obtain that
equation~\thetag{7.12} has precisely~$2n$ roots~\thetag{7.13} in
the~$\veps$-neighbourhood, $\veps>0$, of~$(1,0)$.
\enddemo

\example{Example} For $n=1$ equation~\thetag{7.14} takes the form
$\lambda^2=2$. For $n=2$~it takes the form
$$
\gather
\lambda^4-12\lambda^2+12=0,
\\
\lambda_1^\pm=\pm\sqrt{6+2\sqrt6}\,, \qquad
\lambda_2^\pm=\pm\sqrt{6-2\sqrt6}\,.
\endgather
$$
For $n=3$ it takes the form
$$
\gather
\lambda^6-30\lambda^4+180\lambda^2-120 = 0,
\\
\lambda_1^\pm\approx\pm0{,}87, \qquad \lambda_2^\pm\approx\pm2{,}67,
\qquad \lambda_3^\pm\approx\pm4{,}70.
\endgather
$$
\endexample

\proclaim{Assertion 7.1}  If\/ $\pfi(t)$~is a solution of
equation~\rom{(7.1)} such that\/~\thetag{5.4} holds, then
$$
\alignat1
&\frac1{\sqrt{\pi x}}\INT\pfi^{2q}(\tau)
\exp\biggl\{-\frac{(t-\tau)^2}{x}\biggr\}\,d\tau
\\
&\qquad \le
x^{\frac{q}{2q-1}}(1+x)^{-\frac1{4q-2}}\sqrt{\frac{2q-1}{2qx-x-1}}\,
\tag7.25
\endalignat
$$
for all\/ $x>1/(2q-1)$.
\endproclaim

\demo{Proof} Denoting the left-hand side of inequality~\thetag{7.25}
by~$J(x,t)$, using the boundary conditions~\thetag{5.2}, the properties of
solutions of the heat equation and H\"older's inequality, we obtain the
following chain of relations for all $x>1/(2q-1)$:
$$
\alignat1
J(x,t)
&\equiv\frac1{\sqrt{\pi x}}\INT\pfi^{2q}(\tau)
\exp\biggl\{-\frac{(t-\tau)^2}{x}\biggr\}\,d\tau
\\
&=\frac1{\sqrt{\pi(1+x)}}\INT\pfi(\tau)
\exp\biggl\{-\frac{(t-\tau)^2}{1+x}\biggr\}\,d\tau
\\
&\le\frac1{\sqrt{\pi(1+x)}}\INT\pfi(\tau)
\exp\biggl\{-\frac{(t-\tau)^2}{2qx}\biggr\}
\exp\biggl\{-\frac{(t-\tau)^2(2qx-x-1)}{(1+x)2qx}\biggr\}\,d\tau
\\
&=\frac1{\sqrt{\pi(1+x)}}\,\bigl[\sqrt{\pi t}J\bigr]^{\frac1{2q}}
\biggl(\INT\exp\biggl\{-\frac{(t-\tau)^2(2qx-x-1)}
{x(1+x)(2q-1)}\biggr\}\,d\tau\biggr)^{1-\frac1{2q}},
\endalignat
$$
whence
$$
J^{1-\frac1{2q}}\le\frac{(\pi x)^{\frac1{4q}}}{\sqrt{\pi x(1+x)}}
\Biggl(\,\sqrt{\frac{\pi
x(1+x)(2q-1)}{2qx-x-1}}\,\,\Biggr)^{1-\frac1{2q}},
$$
which implies that~\thetag{7.25} holds.
\enddemo

\proclaim{Corollary 7.1} For\/ $x=1$ estimate~\thetag{7.25} with\/
$q=2,3,\dots$ takes the form
$$
\frac1{\sqrt\pi}\INT\pfi^{2q}(\tau)\exp\{-(t-\tau)^2\}\,d\tau
\le2^{-\frac1{4q-2}}\sqrt{\frac{2q-1}{2q-2}}\,.
\tag7.26
$$
\endproclaim

\head \S\,8. The case when $p=2$. Solution of the equation
\endhead

Equation~\thetag{1.3} with $p=2$ takes the form
$$
\pfi^2(t)=\frac1{\sqrt\pi}\INT\pfi(\tau)e^{-(t-\tau)^2}\,d\tau.
\tag8.1
$$

Let $\pfi(t)$~be a solution of equation~\thetag{8.1} belonging to~$L_2^1$.
We are going to obtain a formally infinite system of non-linear equations
in the coefficients~$a_n$ occurring in expansion~\thetag{4.3} of~$\pfi^2$
and expansion~\thetag{4.1} of~$\pfi$. We obtain the Taylor series
for~$\pfi(t)$ from~\thetag{4.1}, using expansion~\thetag{2.7} of Hermite
polynomials in powers of~$t$. We have
$$
\alignat1
\pfi^2(t)
&=\sum_{m=0}^\infty\frac{a_m}{2^mm!}\,H_m(t)
\sum_{s=0}^\infty\frac{a_s}{2^ss!}\,H_s(t)
\\
&=\sum\Sb m=0 \\ s=0 \endSb^\infty
\frac{a_ma_s}{2^{m+s}}\sum\Sb k=0 \\ k\equiv m \endSb^m
c_{m,k}t^k\sum\Sb i=0 \\ i\equiv s \endSb^s
c_{s,i}t^i
\\
&=\sum\Sb m=0 \\ s=0 \endSb^\infty
\frac{a_ma_s}{2^{m+s}}\sum_{n=0}^{m+s}t^n
\sum\Sb k\equiv m \\ i\equiv s \endSb^{k+i=n}
c_{m,k}c_{s,i}
\\
&=\sum_{n=0}^\infty t^n\sum_{m+s\ge n}^\infty\frac{a_ma_s}{2^{m+s}}
\sum\Sb k\equiv m \\ i\equiv s \endSb^{k+i=n}
c_{m,k}c_{s,i}
\\
&=\sum_{n=0}^\infty t^n\sum_{k+i=n}\sum\Sb m=k \\ m\equiv k \endSb^\infty
\sum\Sb s=i \\ s\equiv i \endSb^\infty
\frac{a_ma_s}{2^{m+s}}\,c_{m,k}c_{s,i}
\\
&=\sum_{n=0}^\infty t^n\sum_{k+i=n}
\biggl(\,\sum\Sb m=k \\ m\equiv k \endSb^\infty
\frac{a_m}{2^m}c_{m,k}\biggr)
\biggl(\,\sum\Sb s=i \\ s\equiv i \endSb^\infty
\frac{a_s}{2^s}c_{s,i}\biggr).
\tag8.2
\endalignat
$$

Comparing~\thetag{8.2} with~\thetag{4.3}, we obtain the desired system of
non-linear equations in~$a_n$:
$$
a_n=n!\sum_{k+i=n}\biggl(\,\sum\Sb m=k \\ m\equiv k \endSb^\infty
\frac{a_m}{2^m}c_{m,k}\biggr)
\biggl(\,\sum\Sb s=i \\ s\equiv i \endSb^\infty
\frac{a_s}{2^s}c_{s,i}\biggr), \qquad
n=0,1,\dots\,.
\tag8.3
$$

Let us note that the series in~\thetag{8.3} converge, which follows from
the Parseval--Steklov equality~\thetag{4.1} and the Cauchy--Bunyakovskii
inequality.

Let us write the first four equations of system~\thetag{8.3}
($n=0,1,2,3$), using equalities~\thetag{2.9} for~$c_{m,k}$. For $n=0$ we
have
$$
a_0=\biggl(\,\sum_{m=0}^\infty(-1)^m\frac{a_{2m}}{4^mm!}\biggr)^2, \tag8.4
$$
which enables us to obtain the following linear equation in~$a_{2m}$, \
$m\ge1$:
$$
\sum_{m=0}^\infty(-1)^m\frac{a_{2m}}{4^mm!}=\veps\sqrt{a_0}\,, \qquad
\veps=\pm1.
\tag8.5
$$

First we consider the case when $a_0>0$. For $n=1$ we have
$$
a_1=2\biggl(\,\sum_{m=0}^\infty(-1)^m\frac{a_{2m}}{4^mm!}\biggr)
\biggl(\,\sum_{s=0}^\infty (-1)^s\frac{a_{2s+1}}{4^ss!}\biggr). \tag8.6
$$
Using~\thetag{8.5}, we obtain the following linear equation in~$a_{2m+1}$,
\ $m\ge0$:
$$
\sum_{m=0}^\infty(-1)^m\frac{a_{2m+1}}{4^mm!}=\frac{a_1}{2\veps\sqrt{a_0}}\,.
\tag8.7
$$

For $n=2$ we have
$$
\alignat1
a_2
&=2\biggl(\,\sum_{m=0}^\infty(-1)^m\frac{a_{2m+1}}{4^mm!}\biggr)^2
\\
&\qquad -8\biggl(\,\sum_{m=0}^\infty(-1)^m\frac{a_{2m}}{4^mm!}\biggr)
\biggl(\,\sum_{s=1}^\infty(-1)^s\frac{a_{2s}}{4^s(s-1)!}\biggr). \tag8.8
\endalignat
$$
Combining this with~\thetag{8.5} and~\thetag{8.7}, we deduce the following
linear equation in~$a_{2m}$, \ $m\ge1$:
$$
\sum_{m=1}^\infty(-1)^m\frac{a_{2m}}{4^m(m-1)!}
=\frac1{8\veps\sqrt{a_0}}\biggl[\frac{a_1^2}{2a_0}-a_2\biggr].
\tag8.9
$$

Finally, for $n=3$ we have
$$
\alignat1
a_3
&=-8\biggl(\,\sum_{m=0}^\infty(-1)^m\frac{a_{2m}}{4^m m!}\biggr)
\biggl(\,\sum_{s=1}^\infty(-1)^s\frac{a_{2s+1}}{4^s(s-1)!}\biggr)
\\
&\qquad -24\biggl(\,\sum_{m=0}^\infty(-1)^m\frac{a_{2m+1}}{4^mm!}\biggr)
\biggl(\,\sum_{s=1}^\infty(-1)^s\frac{a_{2s}}{4^s(s-1)!}\biggr). \tag8.10
\endalignat
$$
Combining this with~\thetag{8.5},~\thetag{8.7} and~\thetag{8.9}, we deduce
the following linear equation in~$a_{2m+1}$, \ $m\ge1$:
$$
\sum_{m=1}^\infty(-1)^m\frac{a_{2m+1}}{4^m(m-1)!}
=-\frac1{8\veps\sqrt{a_0}}\biggl[a_3
-\frac{3a_1}{2a_0}\biggl(\frac{a_1^2}{\veps\sqrt{a_0}}-a_2 \biggr)\biggr].
\tag8.11
$$

Hence, the non-linear system of equations~\thetag{8.3} for $n=0,1,2,3$ is
decomposed into four linear equations -- equations~\thetag{8.5}
and~\thetag{8.9} in~$a_{2m}$, \ $m\ge1$, and equations~\thetag{8.7}
and~\thetag{8.11} in~$a_{2m+1}$, \ $m\ge1$. System~\thetag{8.3} has a
similar structure for all~$n$.

Now assume that $a_0=0$. By~\thetag{8.5}, we have
$$
\sum_{m=1}^\infty(-1)^m\frac{a_{2m}}{4^mm!}=0.
\tag8.12
$$
It follows from~\thetag{8.6} that $a_1=0$, and~\thetag{8.8} implies that
$$
a_2=2\biggl(\,\sum_{m=0}^\infty(-1)^m\frac{a_{2m+1}}{4^mm!}\biggr)^2\ge0.
\tag8.13
$$

If $a_2=0$, which implies, by~\thetag{8.13}, that
$$
\sum_{m=0}^\infty(-1)^m\frac{a_{2m+1}}{4^mm!}=0,
\tag8.14
$$
then~\thetag{8.10} implies that $a_3=0$, and so on. Finally, we find an
integer~$\sigma\ge1$ such that $a_{2\sigma}>0$, whence $a_{2\sigma+1}=0$
(if the solution is different from the identical zero).

Solving~\thetag{8.3} with $a_0=a_1=\dots=a_{2\sigma-1}=0$, we construct
(as in the case when $a_0>0$) a formal solution of equation~\thetag{8.1},
using formula~\thetag{4.1}.

In the case when $a_2>0$ formula~\thetag{8.13} implies that
$$
\sum_{m=0}^\infty(-1)^m\frac{a_{2m+1}}{4^mm!}=\pm\sqrt{\frac{a_2}2}\,.
\tag8.15
$$
Combining this with~\thetag{8.10}, we obtain that
$$
a_3=\pm24\sqrt{\frac{a_2}2}\sum_{m=1}^\infty(-1)^m\frac{a_{2m}}{4^m(m-1)!}\,.
\tag8.16
$$

To obtain an approximate solution of system~\thetag{8.3}, we have to put
$a_n= \mathbreak a_{n+1}=\dots=0$ in the first~$n$ equations (the
$(n-1)$-approximation). In this way we obtain a system of~$n$~equations
in~$n$ unknowns $a_0,a_1,\dots,a_{n-1}$.

Consider the $n$-approximation, $n=3$, for $a_0>0$. We obtain
from~\thetag{8.5}, \thetag{8.7}, \thetag{8.9} and~\thetag{8.11} the
following system of equations in~$a_0$, $a_1$, $a_2$ and~$a_3$:
$$
\alignat1
a_0
&=\veps\sqrt{a_0}+\frac{a_2}4\,, \qquad
a_0\ge0,
\tag8.17${}_0$
\\
a_1
&=2\veps\sqrt{a_0}\,\biggl(a_1-\frac{a_3}{4}\biggr),
\tag8.17${}_1$
\\
a_2
&=\frac{a_1^2}{2a_0}+2\veps\sqrt{a_0}\,a_2,
\tag8.17${}_2$
\\
a_3
&=2\veps\sqrt{a_0}\,a_3+\frac{3a_1a_2}{\veps\sqrt{a_0}}\,.
\tag8.17${}_3$
\endalignat
$$

Consider the linear system of equations in~$a_1$ and~$a_3$ that consists
of equations~\thetag{8.17${}_1$} and~\thetag{8.17${}_3$}:
$$
\gathered
\bigl(1-\veps\sqrt{a_0}\,\bigr)a_1+\frac12\,\veps\sqrt{a_0}\,a_3=0,
\\
-3\,\frac{a_2a_1}{\veps\sqrt{a_0}}+\bigl(1-2\veps\sqrt{a_0}\,\bigr)a_3=0.
\endgathered
\tag8.18
$$
The determinant of this system is equal to
$$
D=1+4a_0-4\veps\sqrt{a_0}+\frac32\,a_2.
$$
Taking into account~\thetag{8.17${}_0$}, we obtain that
$$
D=1+10a_0-10\veps\sqrt{a_0}\,.
\tag8.19
$$

In a similar way, for $a_0=0$ and~$a_2>0$ we obtain from~\thetag{8.15}
and~\thetag{8.16} the following system of equations in~$a_2$ and~$a_3$:
$$
a_3=\pm\sqrt{8a_2}\,, \qquad
a_3=\pm3\sqrt2\,a_2^{3/2}.
\tag8.20
$$

The following assertion holds for the solutions of systems~\thetag{8.17}
and~\thetag{8.20}.

\proclaim{Assertion 8.1} If\/ $a_0>0$ and\/~$D\ne0$, then

{\rm (a)} $a_0=1$ and\/~$a_1=a_2=a_3=0$, which corresponds to the trivial
solution\/ $\pfi(t)=1$,

{\rm (b)} $a_0=1/4$, \ $a_1=0$, \ $a_2=-1$ \ and \ $a_3=0$ {\rm(}in this
case $\veps=1${\rm)}, which corresponds to the approximate solution
$$
\pfi\approx\frac12\,(1-t^2).
\tag8.21
$$

If\/ $a_0>0$ and\/~$D=0$ {\rm(}in this case\/ $\veps=1${\rm)}, then

{\rm (c)} $a_0=0{,}4000+\sqrt{0{,}15}\approx0{,}7873$, \
$a_1\approx\pm0{,}6984$, \ $a_2=-0{,}4000$ \ and \ $a_3\approx\pm1{,}219$,
which corresponds to the following two approximate solutions:
$$
\alignat1
\pfi(t)
&\approx0{,}7873\pm0{,}6984t-0{,}05000H_2(t)\pm0{,}02540H_3(t)
\\
&=0{,}8873\pm0{,}3936t-0{,}2000t^2\pm0{,}2032t^3.
\tag8.22
\endalignat
$$

If\/ $a_0=0$ {\rm(}in this case\/ $a_1=0${\rm)}, then

{\rm (d)} either\/ $a_2=a_3=0$ or\/ $a_2=2/3$ and\/~$a_3=\pm4/\sqrt3$\,, 
which corresponds either to the trivial solution\/ $\pfi(t)=0$ or to the two
approximate solutions
$$
\pfi(t)\approx\frac1{12}\,H_2(t)\pm\frac1{12\sqrt3}\,H_3(t).
\tag8.23
$$
\endproclaim

\demo{Proof} If $a_0>0$ and \ $D\ne0$, then $a_1=a_3=0$. In this case
equations~\thetag{8.17${}_0$} and~\thetag{8.17${}_2$} give solutions~(a)
and~(b) (and $\veps=1$). If $a_0>0$ \ and \ $D=0$, then $\veps=1$
and~$a_0=0{,}4\pm\sqrt{0{,}15}$\,. It follows from~\thetag{8.17${}_0$}
that $a_2=-0{,}4$. Equations~\thetag{8.17${}_2$} and~\thetag{8.17${}_3$}
imply that $a_0=0{,}4+\sqrt{0{,}15}$, and we have two approximate
solutions~(c) with $a_1\approx\pm0{,}6984$ and~$a_3\approx\pm1{,}219$. If
$a_0=a_1=0$, then~\thetag{8.20} implies that assertion~(d) holds.
\enddemo

\remark{Remark 8.1} The question arises whether the approximate
solutions~\thetag{8.21}--\thetag{8.23} are ``parasite''  or they are the
first terms of unknown solutions belonging to~$L_2^1$. We have verified up
to the terms of order of~$t^4$ the squares of solutions~\thetag{8.22}
coincide with the squares of these solutions
$$
\pfi^2(t)\approx0{,}7873\pm0{,}6984t-0{,}2000t^2\pm0{,}2032t^3 \tag8.24
$$
computed by formula~\thetag{4.3}.
\endremark

\head \S\,9. The case when $p=2$. The boundary-value problem
\endhead

The method of solving equation~\thetag{8.1} described in~\S\,8 can be
applied to problem~\thetag{8.1}, \thetag{1.5}.

Let $\pfi$~be a solution of problem~\thetag{8.1},~\thetag{1.5}. We seek
this solution in the form
$$
\pfi(t)=\pfi_0(t)+e^{-(\alpha^2-1)t^2}\sum_{m=0}^\infty c_mH_m(\alpha t),
\tag9.1
$$
where $\alpha$~is a parameter greater than~$1$ and
$$
\pfi_0(t)=\frac12+\frac12\erf(t), \qquad
\erf(t)=\frac2{\sqrt\pi}\int_0^te^{-x^2}\,dx.
\tag9.2
$$
The right-hand side of~\thetag{9.1} satisfies the boundary
conditions~\thetag{1.5}.

We shall be able to use representation~\thetag{9.1} if we verify that the
system of functions
$$
\chi_n(t)=e^{-(\alpha^2-1)t^2}H_n(t), \qquad
n=0,1,\dots,
$$
is a basis of the separable Hilbert space~$L_2^1$. It is sufficient to
prove that this system is complete, that is, to prove that $f=0$ if~$f\in
L_2^1$ is such that
$$
(f,\chi_n)=\frac1{\sqrt\pi}\INT f(t)e^{-(\alpha^2-1)t^2}H_n(t)\,dt=0,
\qquad n=0,1,\dots\,.
$$
This follows immediately from the fact that the system of Hermite
polynomials is complete in~$L_2^1$.

Now we are going to compute the coefficients $a_n=(\pfi,H_n)$ of the
expansion of~$\pfi$ in Hermite polynomials. We have
$$
\alignat1
e_n
&=(\pfi_0,H_n)=\frac1{\sqrt\pi}\INT\pfi_0(\tau)H_n(\tau)e^{-\tau^2}\,d\tau
\\
&=\frac12\,\delta_{n0}+\frac1{2\sqrt\pi}\INT\erf(\tau)H_n(\tau)e^{-\tau^2}\,d\tau
\\
&=\cases
\dfrac1{2\pi}(-1)^{\frac{n-1}2}2^{\frac n2}\Gamma\biggl(\dfrac n2\biggr),
& n\equiv1,
\\
0,
& n\equiv0, \quad
n\ne0,
\\
\dfrac12\,,\vphantom{\bigg|}
& n=0.
\endcases
\tag9.3
\endalignat
$$
Here we used formula~2.20.10.1 in~\cite{11}, p.~497 of the Russian
version, with $b=c=1$.

Using formula 2.20.16.4 in~\cite{11}, p.~502 of the Russian version, with
$p=\alpha^2$, \ $b=\alpha$, \ and \ $c=1$, we compute the integral
$$
\alignat1
&\frac1{\sqrt\pi}\INT e^{-\alpha^2\tau^2}H_m(\alpha\tau)H_n(\tau)\,d\tau
\\
&\qquad =\cases
n!\,2^m\biggl(\dfrac{n-m}{2}\,!\biggr)^{-1}(1-\alpha^2)^{\frac{n-m}{2}}\alpha^{-n-1},
& m\le n\text{\quad and\quad$m\equiv n$},
\\
0, & m>n \text{\quad or\quad $m\not\equiv n$}.
\endcases
\endalignat
$$
Combining this with~\thetag{9.1} and~\thetag{9.3}, we obtain the following
formula for the Hermite coefficients of~$\pfi$:
$$
a_n=e_n+n!\,\alpha^{-n-1}\sum_{\Sb m=0\\m\equiv n\endSb}^nc_m2^m
\biggl(\frac{n-m}{2}!\biggr)^{-1}(1-\alpha^2)^{\frac{n-m}{2}}.
\tag9.4
$$
In particular,
$$
\gather
a_0=\frac12+\frac{c_0}{\alpha}\,,  \qquad
a_1=\frac1{\sqrt{2\pi}}+\frac{2c_1}{\alpha^2}\,, \qquad
a_2=-2c_0\frac{\alpha^2-1}{\alpha^3}+\frac{8c_2}{\alpha^3}\,,
\\                                                          *
a_3=-\frac1{\sqrt{2\pi}}-12c_1\frac{\alpha^2-1}{\alpha^4}
+\frac{48c_3}{\alpha^4}\,.
\tag9.5
\endgather
$$

Substituting expressions~\thetag{9.4} for $a_n$ into~\thetag{8.3}, we
obtain an infinite system of non-linear equations in the unknowns~$c_m$,
which occur in formula~\thetag{9.1} for~$\pfi$.

Considering the~$3$-approximation, we put in~\thetag{9.5} for
definiteness $a_0=0{,}7873$, \ $a_1=0{,}6984$, \ $a_2=-0{,}4000$ \ and \
$a_3=1{,}219$ (see Assertion~7.1) and obtain the following equations
in~$c_0$, \ $c_1$, \ $c_2$ and~$c_3$:
$$
\gather
0{,}7873=\frac12+\frac{c_0}{\alpha}\,,  \qquad
0{,}6984=\frac1{\sqrt{2\pi}}+\frac{2c_1}{\alpha^2}\,,
\\
-0{,}4000=-2c_0\frac{\alpha^2-1}{\alpha^3}+\frac{8c_2}{\alpha^3}\,, \qquad
1{,}219=-\frac1{\sqrt{2\pi}}-12c_1\frac{\alpha^2-1}{\alpha^4}
+\frac{48c_3}{\alpha^4}\,.
\endgather
$$
Hence,
$$
\gather
c_0=0{,}2873\alpha, \qquad
c_1=0{,}1498\alpha^2,
\\
c_2=0{,}02182\alpha^3-0{,}07182\alpha, \qquad
c_3=0{,}07120\alpha^4-0{,}03746\alpha^2.
\tag9.6
\endgather
$$

Putting~$\alpha^2=1{,}1$ in~\thetag{9.6}, we obtain that
$$
c_0=0{,}3014, \qquad
c_1=0{,}1648, \qquad
c_2=-0{,}05016, \qquad
c_3=0{,}04494.
$$
Substituting these numbers into~\thetag{9.1}, we obtain the following
approximate solution of problem~\thetag{8.1}, \thetag{1.5}:
$$
\alignat1
\pfi(t)
&\approx\frac12+\frac12\,\erf(t)+e^{-0,1t^2}\bigl(0{,}3014+0{,}1648H_1\bigl(\sqrt{1{,}1}\,t\bigr)
\\
&\qquad
-0{,}05016H_2\bigl(\sqrt{1{,}1}\,t\bigr)+0{,}04494H_3\bigl(\sqrt{1{,}1}\,t\bigr)\bigr)
\\
&\quad
=\frac12+\frac12\,\erf(t)+e^{-0{,}1t^2}\bigl(0{,}4017-0{,}2200t-0{,}2207t^2
+0{,}4149t^3\bigr).
\tag9.7
\endalignat
$$
The second solution can be obtained likewise if we consider~$a_1$
and~$a_3$ with minus sign.

\remark{Remark 9.1} Our choice $\alpha=\sqrt{1{,}1}\approx1{,}049$ was an
arbitrary one. The method does not depend on~$\alpha>1$ as long as it
ranges between reasonable limits. In order to find the optimal value
of~$\alpha$ one should perform supplementary calculations.
\endremark

\head \S\,10. The case when $p$ is odd ($p=2q+1$)
\endhead

Equation~\thetag{1.3} with $p=2q+1$ takes the form
$$
\pfi^{2q+1}(t)=\frac1{\sqrt\pi}\INT\pfi(\tau)e^{-(t-\tau)^2}\,d\tau, \qquad
q=1,2,\dots\,.
\tag10.1
$$

If $\pfi(t)$ is a solution of equation~\thetag{10.1}, then~$-\pfi(t)$, \
$\pfi(-t)$ and~$\pfi(t+t_0)$ with any~$t_0$ also are solutions.

We can find solutions of equation~\thetag{10.1}, using the method of
expanding in Hermite polynomials described in~\S\,7--9, but the
calculations involved would be bulky even in the case when~$p=3$. It is
also possible to solve problem~\thetag{10.1}, \thetag{1.6}, using the
following substitution similar to~\thetag{9.1}:
$$
\pfi(t)=\erf(t)+e^{-(\alpha^2-1)t^2}\sum_{m=0}^\infty c_mH_m(\alpha t).
$$

In~\cite{9} it was proved that problem~\thetag{10.1}, \thetag{1.6} has a
continuous odd solution real-analytic for $t\ne0$ that has precisely one
real zero at~$t=0$, and
$$
\pfi(t)=(a_1t)^{\frac1{2q+1}}[1+O(|t|)], \qquad
t\to0,
\tag10.2
$$
where
$$
a_1=\frac4{\sqrt\pi}\int_0^\infty\pfi(\tau)e^{-\tau^2}\tau\,d\tau>0.
\tag10.3
$$

As in the case when~$p$ is even (see~\S\,7), the following theorem holds.

\proclaim{Theorem 10.1} If\/ $\pfi(t)$~is a solution of
problem~\thetag{10.1}, \thetag{1.6}, then it is continuous,
$\pfi^{2q+1}(t)$ has finitely many zeros\/~$t_k$ of finite
multiplicity\/~$\sigma_k$, \ $k=1,2,\dots,l$, \ $\sum_{k=1}^l\sigma_k$~is an
odd number, and
$$
\pfi(t)=\biggl[\frac{a_{\sigma_k}}{(\sigma_k)!}\biggr]^{\frac{1}{2q+1}}
(t-t_k)^{\frac{\sigma_k}{2q+1}}[1+O(|t-t_k|)], \qquad
t\to t_k,
\tag10.4
$$
where
$$
\frac{2^{\sigma_k}}{\sqrt\pi}\INT\pfi(\tau)(\tau-t_k)^ne^{-(t_k-\tau)^2}\,d\tau
=\cases
a_{\sigma_k},
& n=\sigma_k,
\\
0,
& n=0,1,\dots,\sigma_k-1.
\endcases
\tag10.5
$$
The number of sign changes of\/~$\pfi(t)$ is odd and coincides with that
of\/~$\pfi^{2q+1}(t)$. This number is less than or equal to\/~\,$l$ and
greater than or equal to\/~$\max_{1\le k\le l}\sigma_k$.
\endproclaim

I am grateful to I.~V.~Volovich, A.~K.~Gushchin and~V.~P.~Mikhailov for
useful discussion of the problems treated in this paper.


\enddocument